\documentclass[sigconf, nonacm]{acmart}

\usepackage{amsmath,amsfonts}
\usepackage{graphicx}
\usepackage{textcomp}
\usepackage{xcolor}
\usepackage{enumitem}
\usepackage{tikz}
\usepackage{booktabs}
\usepackage{hyperref}
\usepackage{tabularx}
\usepackage[labelformat=simple]{subfig}
\usetikzlibrary{arrows.meta, positioning, shapes.geometric, matrix}

\newcommand{\systemname}{SafeKV}

\newcommand{\circnum}[1]{%
  \tikz[baseline=(char.base)]{
    \node[shape=circle, fill=black, inner sep=0.7pt] (char)
      {\color{white}\scriptsize #1};
  }%
}

\begin{document}
% \title{\systemname: System-Level Co-Design of Privacy Enforcement and KV-Cache Management for LLM Serving}
\title{Selective KV-Cache Sharing to Mitigate Timing Side-Channels in LLM Inference}

\author{Kexin Chu}
\authornote{These authors contributed equally to this work.}
\affiliation{%
  \institution{University of Connecticut}
  \city{Storrs}
  \state{CT, USA}
}
\email{kexin.chu@uconn.edu}

\author{Zecheng Lin}
\authornotemark[1]
\affiliation{%
  \institution{Independent}
}
\email{zc-lin24@mails.tsinghua.edu.cn}

\author{Dawei Xiang}
\affiliation{%
  \institution{University of Connecticut}
  \city{Storrs}
  \state{CT, USA}
}
\email{ieb24002@uconn.edu}

\author{Zixu Shen}
\affiliation{%
  \institution{University of Connecticut}
  \city{Storrs}
  \state{CT, USA}
}
\email{qzt24001@uconn.edu}

\author{Jianchang Su}
\affiliation{%
  \institution{University of Connecticut}
  \city{Storrs}
  \state{CT, USA}
}
\email{jianchang.su@uconn.edu}

\author{Cheng Chu}
\affiliation{%
  \institution{Indiana University Bloomington}
  \city{Bloomington}
  \state{IN, USA}
}
\email{chu6@iu.edu}

\author{Yiwei Yang}
\affiliation{%
  \institution{UC Santa Cruz}
  \city{Santa Cruz}
  \state{CA, USA}
}
\email{yyang363@ucsc.edu}

\author{Wenhui Zhang}
\affiliation{%
  \institution{Independent}
}
\email{wenhuizhang.psu@gmail.com}

\author{Wenfei Wu}
\affiliation{%
  \institution{Peking University}
  \city{Beijing}
  \state{China}
}
\email{wenfeiwu@pku.edu.cn}

\author{Wei Zhang}
\authornote{Corresponding author.}
\affiliation{%
  \institution{University of Connecticut}
  \city{Storrs}
  \state{CT, USA}
}
\email{wei.13.zhang@uconn.edu}

\vskip 0.3in

\begin{abstract}

Global KV-cache sharing is an effective optimization for accelerating large language model (LLM) inference, yet it introduces an API-visible timing side channel that lets adversaries infer sensitive user inputs from shared entries, leading to cross-tenant privacy risks. To address this problem, we introduce \systemname\ (Secure and Flexible KV-cache Sharing), a system-level co-design of privacy enforcement and KV-cache management. \systemname\ integrates lightweight detection and isolation directly into the serving runtime to eliminate cross-tenant reuse of sensitive KV-cache blocks under our threat model, while recovering most of the performance benefits of global sharing. Our key contributions are: (1) a three-tier asynchronous detection pipeline that decouples privacy classification from inference and supports streaming workloads, (2) a unified radix-tree–based memory manager with path compression and sensitivity-aware eviction for scalable selective isolation, and (3) an RDR-guided (Reuse Diversity Ratio) runtime safeguard that detects and bounds residual leakage. On large LLM backends, \systemname\ reduces the time-to-first-token (TTFT) overhead compared to full isolation by up to $40.58\%$ and raises throughput by up to $2.66\times$. Overall, \systemname\ restores the efficiency of KV reuse while enforcing strong, practical privacy for multi-tenant LLM inference.

\end{abstract}
\maketitle

% \pagestyle{\vldbpagestyle}
% \begingroup\small\noindent\raggedright\textbf{PVLDB Reference Format:}\\
% \vldbauthors. \vldbtitle. PVLDB, \vldbvolume(\vldbissue): \vldbpages, \vldbyear.\\
% \href{https://doi.org/\vldbdoi}{doi:\vldbdoi}
% \endgroup
% \begingroup
% \renewcommand\thefootnote{}\footnote{\noindent
% This work is licensed under the Creative Commons BY-NC-ND 4.0 International License. Visit \url{https://creativecommons.org/licenses/by-nc-nd/4.0/} to view a copy of this license. For any use beyond those covered by this license, obtain permission by emailing \href{mailto:info@vldb.org}{info@vldb.org}. Copyright is held by the owner/author(s). Publication rights licensed to the VLDB Endowment. \\
% \raggedright Proceedings of the VLDB Endowment, Vol. \vldbvolume, No. \vldbissue\ %
% ISSN 2150-8097. \\
% \href{https://doi.org/\vldbdoi}{doi:\vldbdoi} \\
% }\addtocounter{footnote}{-1}\endgroup
% %%% VLDB block end %%%

% %%% do not modify the following VLDB block %%
% %%% VLDB block start %%%
% \ifdefempty{\vldbavailabilityurl}{}{
% \vspace{.3cm}
% \begingroup\small\noindent\raggedright\textbf{PVLDB Artifact Availability:}\\
% The source code, data, and/or other artifacts have been made available at \url{https://github.com/kexinchu/InferShield}. This repository hosts the SafeKV prototype integrated into SGLang.
% \endgroup

\section{Introduction}
\label{sec:intro}

Large language models (LLMs) now drive applications from dialogue to complex reasoning. To meet time-sensitive inference demands, key–value (KV) caching stores intermediate attention states (“keys” and “values”) to eliminate redundant computation for sequential or similar prompts, thereby accelerating generation~\cite{zhao2024buzz}. This efficiency gain is amplified through KV-cache sharing across multiple requests. As a result, KV-cache sharing is widely used to boost throughput and reduce response latency in large-scale, multi-user LLM deployments~\cite{li2024survey, zheng2024batchllm,chu2025mcam}.

\begin{figure}[t]
    \centering
    \includegraphics[width=0.99\columnwidth]{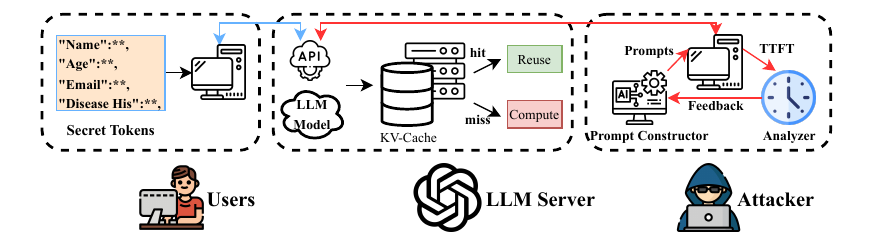}
    \caption{Attack Overview.}
    \label{fig:threat_model}
    % \vspace{-1.0em}
\end{figure}

However, KV-cache sharing raises serious privacy and security concerns in shared or multi-tenant deployments. Specifically, sharing KV entries across mutually untrusted users can leak information. As shown in~\autoref{fig:threat_model}, an adversary can infer cache hits by issuing carefully crafted queries and measuring response latencies. These timing differences reveal whether a prefix has been cached before, which can expose other users' query patterns. With enough probes, cache probing attacks can recover partial or even complete user inputs~\cite{song2024early, wu2025prompt, zheng2024inputsnatch, wang2025sok}.

These side-channel attacks are concerning for two reasons. First, they require no special privileges: the attacker simply interacts with the LLM through its standard API and can blend in with normal traffic. Second, prompts often contain sensitive data, such as medical questions, financial details, or private instructions to LLM agents. Prior work reports high success rates under practical settings: in some cases, success approaches or exceeds 89\% with query costs on the order of $\sim$10$^2$ requests per token (~\autoref{tab:prior_kv_attacks}). Given the low barrier to attack and the severity of the exposed data,  KV-cache timing leakage is a pressing threat for real deployments.

\begin{table}[t]
  \centering
  \small
  \caption{Reported effectiveness and query cost of recent KV-cache timing side-channel attacks.}
  \vspace{-0.5em}
  \label{tab:prior_kv_attacks}
  \setlength{\tabcolsep}{4pt} % tighten column padding if needed
  \begin{tabularx}{\columnwidth}{l X c}
    \toprule
    \textbf{Work} & \textbf{Avg. success rate} & \textbf{Avg. attack cost} \\
    \midrule
    \textsc{PromptPeek}~\cite{wu2025prompt} &
    99\% / 98\% / 95\% (input / template / whole-prompt) &
    148--306 req/token \\
    \textsc{EarlyBird}~\cite{song2024early} &
    89\% &
    112.72 req/token \\
    \textsc{InputSnatch}~\cite{zheng2024inputsnatch} &
    43.33\%--100\% across 13 legal domains &
    N/R\textsuperscript{$\dagger$} \\
    \bottomrule
  \end{tabularx}

  \raggedright
  \footnotesize
  \textsuperscript{$\dagger$}\,N/R: not reported in the original paper under a comparable ``req/token'' measure.
  \vspace{-1.0ex}
\end{table}

\begin{figure}[t]
    \centering
    \includegraphics[width=0.9\linewidth]{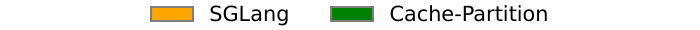}\\[-0.7em]
    \subfloat[Llama-2-13B]{
        \includegraphics[width=0.47\linewidth]{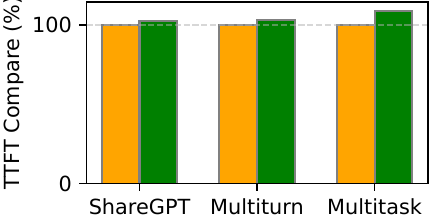}
    }
    \subfloat[Llama-2-70B]{
        \includegraphics[width=0.47\linewidth]{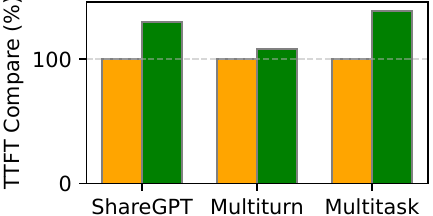}
    }
    \caption{Normalized performance of TTFT between \textit{SGLang (global-sharing)} and \textit{Cache-Partition (isolated-per-user)}.}
\label{fig:compare}
\vspace{-1.0em}
\end{figure}

A straightforward mitigation is timing obfuscation, such as TTFT padding or injected jitter. This approach is poorly aligned with how KV-cache timing attacks work in practice. The attacker does not need exact latency values; repeated probing only needs a reliable statistical separation between the TTFT distributions of cache hits and misses. Under real workloads, cache-miss TTFT can be heavy-tailed, so defeating common two-sample tests may require padding hit responses toward high quantiles of the miss distribution (or injecting jitter of similar magnitude). This directly reduces the user-visible gains of prefix reuse and can inflate P95/P99 TTFT. For example, prior work reports that context caching can reduce TTFT by up to 73\%~\cite{yanglearned}; padding hit responses toward the miss tail would remove most of this gain. A more KV-cache--aware defense is \emph{per-user cache isolation}~\cite{pang2024cache}. Unlike padding or jitter, isolation targets the root cause of cross-tenant leakage by preventing one user from observing another user's cache hits. At the same time, it preserves KV-cache reuse \emph{within} each user, which matches the access pattern of many real applications (e.g., multi-turn sessions and repeated prefixes from the same tenant). This makes isolation a more practical baseline for KV-cache settings. However, it still forgoes the benefits of \emph{cross-tenant} reuse that global caching relies on. In our measurements, isolating caches inflates TTFT by about 38\% and reduces throughput (~\autoref{fig:compare}). This leaves a practical gap: the system should isolate sensitive KV blocks while still enabling reuse on prefixes that are safe to share.

This paper introduces \textbf{\systemname, a high-performance KV-cache manager that integrates privacy enforcement directly into the serving runtime}. In contrast to prior approaches, \systemname\ co-designs privacy detection with cache indexing and memory management. It runs privacy detection asynchronously using lightweight queues, so classification does not stall inference. By co-locating privacy metadata with cache control logic, \systemname\ preserves the low latency and high reuse benefits of global caching while providing system-level protection against timing side channels.

This work makes the following systems contributions:
\begin{itemize}[leftmargin=1.5em, itemsep=1pt, topsep=2pt]
    \item \textbf{Asynchronous Hybrid Detection.} We design a multi-tier detection pipeline that combines rule-based matching, a general privacy detector, and context-aware validation. The pipeline runs asynchronously and is optimized for high-throughput LLM serving.
    \item \textbf{Defense Under Misclassification.} We add runtime safeguards that limit the impact of imperfect detection, including private-by-default insertion and fallback handling for suspicious access. These safeguards keep sensitive data confined without stopping inference.
    \item \textbf{Privacy-Aware Cache Management.} We implement a unified radix-tree cache index that supports both private and shareable entries, with compression-aware indexing and multi-tier memory coordination. The design enables fast lookup and high reuse with small overhead under multi-tenant workloads.
    \item \textbf{Implementation and End-to-End Evaluation.} We implement \systemname\ and evaluate it against state-of-the-art KV-cache isolation baselines: Cache-Partition and Shared-System-Prompt. Across our workloads, \systemname\ improves throughput by up to $2.66\times$ while enforcing cross-tenant privacy constraints.
\end{itemize}

% In combination, these components demonstrate that runtime privacy enforcement can be architected as a first-class system primitive, rather than a post hoc security patch, enabling both efficient and safe cache reuse in LLM serving. This work focuses on integrating practical side-channel defenses into the serving stack, rather than redefining cryptographic guarantees.

\section{Background \& Motivation}
\label{sec:background}

\subsection{LLM Inference and KV-cache Sharing}

\begin{figure}[t]
  \centering
  \includegraphics[width=0.99\columnwidth]{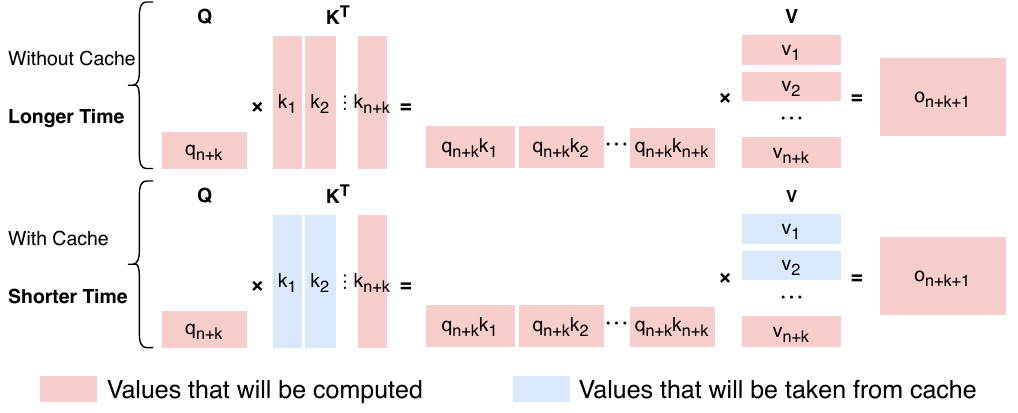}
  \caption{Comparison of self-attention computation mechanisms. The traditional approach (upper) performs full recomputation for each token, while the KV-cache (lower) reuses stored key-value vectors to accelerate inference. The KV-cache reduces the computational complexity per decoding step from $O(n^2)$ to $O(n)$.}
  \label{fig:kv_reuse_overview}
  \vspace{-0.5em}
\end{figure}

Transformer-based LLMs model contextual dependencies through scaled dot-product attention over Query (Q), Key (K), and Value (V) vectors~\cite{lin2022survey,khan2022transformers,wolf2020transformers}. Inference consists of a \emph{prefill} stage, which processes the full prompt to produce the initial per-token K/V vectors and the first output token, followed by a \emph{decoding} stage that generates subsequent tokens one by one. During decoding, the model needs attention over all previous tokens. Without caching, the server would recompute the K/V vectors for the prefix repeatedly. With KV caching, previously computed K/V vectors are stored and reused, and only the new token's K/V vectors are computed at each step (~\autoref{fig:kv_reuse_overview}). This reuse reduces the computational complexity \emph{per decoding step} from $O(n^2)$ to $O(n)$, improving efficiency for long contexts and multi-turn interactions~\cite{ho2024block}.

To improve inference efficiency, LLM serving systems widely enable \emph{KV-cache sharing}, allowing the reuse of previously computed key–value embeddings across requests with overlapping prompt prefixes, as summarized in~\autoref{tab:cache-comparison}. Retention policies vary across providers, but they all aim to exploit prefix overlap for higher throughput. \systemname\ builds upon this standard optimization, addressing how to retain its performance advantages while preventing cross-tenant information leakage through shared caches.

\begin{table}[t]
\centering
\caption{Comparison of cache mechanisms for LLM APIs}
\vspace{-0.5em}
\label{tab:cache-comparison}
\resizebox{\columnwidth}{!}{
  \begin{tabular}{lccc}
    \toprule
    \textbf{Vendor/Framework} & \textbf{Streaming} & \textbf{Caching Mechanism} & \textbf{Cache Lifetime} \\
    \midrule
    OpenAI~\cite{openai2024promptcaching} & Y & Prefix Caching & 5--10 minutes \\
    DeepSeek~\cite{deepseek2024promptcaching} & Y & Prefix Caching & Hours to days \\
    Anthropic Claude~\cite{anthropic2024promptcaching} & Y & Prefix Caching & 5 minutes \\
    Google Gemini~\cite{deepmind-gimini} & Y & Prefix Caching & Default 1 hour \\
    MoonShot Kimi~\cite{qin2025mooncake} & Y & Prefix Caching & Customization \\
    vLLM~\cite{kwon2023efficient} & Y & Prefix Caching & Customization \\
    SGLang~\cite{zheng2024sglang} & Y & Prefix Caching & Customization \\
    \bottomrule
  \end{tabular}
}
\vspace{-1.5em}
\end{table}

\subsection{Timing Side-Channel Attack}

While shared KV caching significantly improves inference efficiency, it also introduces a software-level timing side channel. By issuing crafted queries and measuring TTFT latency, an attacker can distinguish cache hits from misses and thereby infer whether specific prefix tokens were previously used by other users. Recent studies~\cite{wu2025prompt,zhang2024time,soleimani2025wiretapping,zheng2024inputsnatch,song2024early,chu2025safekv} show that such attacks can be executed entirely through black-box APIs: subtle latency variations allow the attacker to iteratively test candidate prefixes and reconstruct parts of a victim’s prompt. 

This leakage extends beyond exact-prefix reuse. Even semantic or partial-match caching can create measurable latency gaps, since queries similar to previously served inputs are processed faster~\cite{bang2023gptcache}. Empirical analyses on both commercial and open-source serving stacks reveal a persistent tension between performance-driven cache reuse and cross-tenant privacy in real-world deployments.

\begin{table}[h]
  \centering
  \caption{Personal Information Counts in C4 and Pile.}
  \vspace{-0.5em}
  \resizebox{\columnwidth}{!}{
  \label{tab:info_counts}
  \begin{tabular}{lrr}
    \toprule
    Personal Information Type & C4           & Pile          \\
    \midrule
    User Name                                 & 1,444,683,066 & 3,273,163,949 \\
    Phone Number                              &    19,592,273 &    23,191,595 \\
    Email Number                              &     9,056,833 &    13,336,793 \\
    US Bank Number                            &     7,139,838 &    69,763,678 \\
    Credit Card Number                        &        61,405 &       741,815 \\
    US SSN                                    &     2,352,339 &    12,541,022 \\
    IP Address                                &     1,890,090 &    14,975,663 \\
    \midrule
    Total                                     & 1,484,780,621 & 3,407,722,116 \\
    \bottomrule
  \end{tabular}
  }
  \vspace{-1.0em}
\end{table}

\subsection{Motivation: Privacy Risk and Isolation Cost}
\vspace{-1.0 em}
\noindent\paragraph{\textbf{The Risk of KV-cache Leakage.}}
To evaluate the privacy risk of KV-cache sharing, we examined two widely used corpora, C4~\cite{2019t5} and Pile~\cite{kandpal2025commonpilev018tb}. As shown in~\autoref{tab:info_counts}, both contain substantial amounts of personally identifiable information (PII), including usernames, phone numbers, credit card numbers, and U.S.\ Social Security Numbers. When cache entries embedding such prefixes are shared across tenants, attackers can exploit latency-based probes to infer their presence and gradually reconstruct sensitive content. Preventing unintended reuse of these tokens is therefore essential in multi-tenant serving.

\noindent\paragraph{\textbf{Performance Impact of Full Isolation.}}
A natural defense is per-user cache isolation, which prevents cross-tenant leakage but eliminates shared reuse~\cite{pang2024cache}. This approach forces identical prefixes to be recomputed and stored separately across HBM, DRAM, and SSD tiers, shrinking effective capacity and increasing TTFT~\cite{lucas2025kv,zeng2025mpcache}. Our motivation experiments (\autoref{fig:compare}) show that Cache-Partition increases TTFT by 2.3\%–8.9\% for Llama-2-13B~\cite{llama2-13b} and by 8.3\%–38.9\% for Llama-2-70B~\cite{llama2-70b} under realistic cross-user overlap. 

\noindent\paragraph{\textbf{\systemname: Bridging Privacy and Performance.}}
To address this trade-off, we propose \textbf{\systemname}, a selective KV-cache sharing framework that classifies cache entries by privacy sensitivity. Non-private entries are safely reused across requests, while sensitive ones remain confined to per-user space. This policy preserves most latency and throughput benefits of shared caching while closing the primary channel for timing-based leakage.

\subsection{Challenges}

However, designing such a system raises three challenges:
\begin{itemize}[leftmargin=1.5em, itemsep=1pt, topsep=2pt]
    \item \textbf{Challenge 1: Accurate and Efficient Privacy Classification.} The system must distinguish sensitive from non-sensitive KV entries with high precision and minimal latency overhead. Rule-based filters are efficient but suffer from limited recall, whereas deep models improve accuracy at the cost of inference delay.
    \item \textbf{Challenge 2: Risk Mitigation under Imperfect Detection.} Even rare misclassifications can cause privacy breaches when sensitive content is mistakenly shared. The system must detect and contain such events dynamically, ensuring safety without interrupting inference execution.
    \item \textbf{Challenge 3: Scalable Cache Lifecycle Management.} Organizing private and shared caches to maximize reuse while maintaining isolation requires careful data placement and indexing. The design must enable fast prefix matching, reduce redundancy, and prevent structural fragmentation under dynamic workloads.
\end{itemize}

\begin{table}[t]
  \centering
  \caption{Intra-session and Inter-session KV-cache reuse rates across different datasets.}
  \vspace{-0.5em}
  \resizebox{\columnwidth}{!}{
  \begin{tabular}{lcc}
    \toprule
    \textbf{Dataset} & \textbf{Intra-User Reuse} & \textbf{Inter-User Reuse} \\
    \midrule
    ShareGPT V3~\cite{sharegpt-chat} & 7.06\% & 25.49\% \\
    Multiturn Chat~\cite{multiturn_chat} & 31.47\% & 9.45\% \\
    Prompt Multitasks~\cite{gallego2024configurable} & 0.0\% & 63.10\% \\
    \bottomrule
  \end{tabular}
  }
\label{tab:reuse-rates}
\vspace{-1.0em}
\end{table}

\section{Threat Model}
\label{sec:threat_model}

\systemname\ targets multi-tenant LLM inference backends where many users share the same serving fleet and the service may reuse KV-cache entries across sessions. We consider an adaptive black-box adversary who can create arbitrary accounts, issue strategically timed queries, and run probes concurrently or in batches. They can coordinate across accounts, replay prompts with controlled prefixes, and adaptively refine probes to test whether a specific prefix or span has been cached by another tenant. The adversary can observe response timing, including TTFT and per-token latency. The adversary cannot read other users’ plaintext prompts, model weights, or KV contents, cannot execute code on serving nodes, and cannot modify the tokenizer or model parameters. We assume the adversary does not break cryptographic primitives, bypass standard authentication, or tamper with the timing source provided by the operating system or hypervisor.

We trust secure transport (TLS) and standard service authentication, and we assume the model and tokenizer artifacts used by the service are correct. We do not trust the multi-tenant workload mix, co-scheduling decisions, or naive KV-sharing policies to preserve privacy by default. Any cross-tenant reuse must be justified and enforced by \systemname. \systemname\ enforces this control through multi-tier privacy detection, creator-bounded access checks, and a runtime safeguard. The safeguard monitors reuse behavior and updates the sharing status of the corresponding KV-cache blocks when it detects suspicious traffic. Our scope is software-visible timing leakage induced by cross-tenant KV-cache reuse in multi-tenant inference. We do not address attacks that require privileged internal access, denial of service, or side channels outside the API timing surface.

\section{Design: \systemname}
\label{sec:design}

To address the tension between privacy protection and high performance cache reuse in large language model serving, we designed and built \systemname, a system-level KV-cache management framework that embeds lightweight privacy classification and isolation directly into the KV-cache memory hierarchy. \systemname\ treats privacy as a first-class design constraint rather than a separate detection workload, and preserves the throughput and latency benefits of shared KV-cache while keeping cross-tenant leakage within measurable bounds. To guide the design, \systemname\ follows three principles:
\begin{itemize}[leftmargin=1.5em, itemsep=1pt, topsep=2pt]
    \item \textbf{Privacy-Aware Reuse.} We encode cache visibility (shareable vs. private) directly in the core KV-cache index to enable cross-tenant reuse when safe while preventing reuse of sensitive blocks across user boundaries.
    \item \textbf{Latency Transparency.} We keep privacy classification off the critical inference path by scheduling classification asynchronously and by defaulting to private until verified shareable, to avoid increasing TTFT or tail latency.
    \item \textbf{Scalable Lifecycle Management.} We provide consistent cache operations (allocation, lookup, eviction) under large-scale multi-tenant workloads with mixed sensitivity by enforcing a single unified lifecycle policy over all blocks.
\end{itemize}

\subsection{System Overview}

\begin{figure}[t]
    \centering
    \includegraphics[width=0.98\columnwidth]{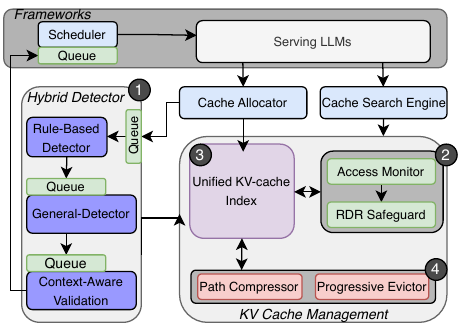}
    \caption{The Architecture Overview of \systemname.}
    \label{fig:arch}
    \vspace{-0.5em}
\end{figure}

The high-level architecture of \systemname, shown in~\autoref{fig:arch}, augments a running LLM serving stack with: (i) an asynchronous privacy detection pipeline, and (ii) a unified KV-cache memory manager. These two subsystems work together to enable safe cache reuse across tenants.

\paragraph{\textbf{Asynchronous Detection Pipeline.}}
To avoid inference stalls, \systemname\ performs privacy classification off the critical serving path via a multi-tier asynchronous pipeline. When a KV block is first generated, the system conservatively marks it as \emph{private} and admits it into the cache under isolation-by-default; the block is upgraded to \emph{shareable} only after background verification completes. This pipeline is built around a \textbf{Hybrid Detector} (\circnum{1}) that increases semantic strength while controlling overhead: (i) Tier-1 rule-based filters quickly catch obvious sensitive patterns and skip clearly safe cases, (ii) Tier-2 lightweight transformer classifiers provide high-throughput semantic screening, and (iii) Tier-3 large-model validation is invoked only for low-confidence cases. To ensure that residual errors do not turn into visible reuse, \systemname\ adds a runtime fallback mechanism (\circnum{2}): a \textbf{KV-Cache Access Monitor} records per-entry reuse statistics during cache lookups, and the \textbf{RDR Safeguard} uses these signals to flag suspicious reuse behavior (e.g., sudden probe-like bursts that deviate from normal tenant traffic). When triggered, the safeguard immediately demotes the entry back to private. This design addresses \textbf{Challenges 1 and 2}: it keeps detection off the serving path to protect latency, and it provides a conservative feedback loop that limits the impact of misclassification.

\paragraph{\textbf{Unified Memory Manager.}} 
\systemname\ realizes privacy-aware cache management through a \textbf{unified radix-tree--based index} (\circnum{3}) that tracks both \emph{shareable} and \emph{private} KV-cache entries under a single set of allocation, lookup, and eviction rules. Instead of maintaining separate structures, the unified index stores visibility as first-class metadata on each cached prefix, allowing the \emph{Cache Search Engine} to check \texttt{private\_tag} at lookup time while making reuse decisions. On top of this unified structure, \systemname\ implements \textbf{path compression} (\circnum{4}) to reduce lookup depth and pointer chasing, enabling fast prefix matching for prompt reuse. It also applies a sensitivity-aware \textbf{progressive evictor} (\circnum{4}) to coordinate multi-tier reclamation while accounting for each entry's visibility and reuse value, so private content is not evicted unfairly and shareable regions are not filled with low-value blocks. Once background detection completes, the system updates an entry's visibility bit atomically, so cache operations always observe consistent access semantics. These two components address \textbf{Challenge 3} by optimizing cache lifecycle management, reducing redundant copies caused by selective isolation, and preserving reuse efficiency without sacrificing correctness.

\subsection{Asynchronous Detection Pipeline}
\label{sec:detection}

First, we focus on the privacy detection subsystem, which classifies each KV-cache block as either private or shareable and enforces runtime safeguards against potential misclassifications. The design treats privacy protection as a system constraint rather than a standalone ML task: detection must integrate tightly with the KV-cache management, preserve inference efficiency, and remain robust under imperfect predictions. To achieve these goals, \systemname\ integrates a \textbf{three-tier hybrid detection pipeline} (\autoref{sec:det-tiered}), a \textbf{runtime fallback mechanism} (\autoref{sec:det-fallback}), and an \textbf{asynchronous scheduler} (\autoref{sec:det-async}).

\subsubsection{Three-Tier Hybrid Detection Strategy}
\label{sec:det-tiered}

To enable \textbf{accurate and efficient privacy classification} for KV-cache blocks, \systemname\ uses a three-tier detection pipeline that filters inputs with \emph{increasing cost and increasing semantic strength}. The key idea is to make most decisions with low overhead, and reserve expensive reasoning for rare corner cases. Specifically, the pipeline proceeds as follows: (i) Tier-1 handles common, explicit identifiers with near-zero cost using deterministic pattern rules; (ii) Tier-2 covers implicit or loosely formatted privacy signals using a small semantic model with low latency; and (iii) Tier-3 resolves the remaining hard cases by validating sensitivity under full conversational context. The concrete design of each tier is described as follow:

\paragraph{\textbf{Tier 1: Rule-Based Pattern Matching}}

\begin{table}[t]
\centering
\caption{Taxonomy of User-Related Data Categories}
\vspace{-0.5em}
\resizebox{\columnwidth}{!}{%
\begin{tabular}{lcc}
\toprule
\textbf{Category} & \textbf{Type} & \textbf{Examples} \\
\midrule
Privacy & General Information & Nickname, avatar, signature \\
(Personal) & Basic Information & Third-party account information \\
~ & Identity Information & ID card, passport, driver’s license, SSN \\
~ & Location Information & Country, region \\
~ & Biometric Identification & Fingerprints, face, voiceprint, iris, gene info \\
~ & System/Network Identification & UserID, IP, Cookie, RFID, password, certs \\
Device & Software Device Information & Android ID, IDFA, IDFV, OS, region \\
~ & Hardware Device Information & MAC, IMEI, GUID, serial number \\
Profile & Cultural \& Social Info & Job, education, qualification certificates \\
~ & Financial Info & Bank account, property, loan records \\
~ & Social Info & Likes/follows, contacts, collections \\
~ & Service Content Info & Browsing/purchase/download records \\
Behavior & Service Log Info & Login, behavior, purchase logs \\
\bottomrule
\end{tabular}
}
\label{tab:user_data_3col}
% \vspace{-1.0em}
\end{table}

Tier-1 focuses on identifiers with stable, easy-to-match formats. It applies deterministic matching to the raw text associated with each KV block using two rule types: (i) regular expressions for variable-length patterns (e.g., emails, phone numbers, and ID strings) and (ii) token blacklists for fixed keywords (e.g., internal project codes or organization-specific keys). The rule set is configurable and supports hot reloading, so operators can add or update patterns without restarting the service.

\autoref{tab:user_data_3col} lists the main categories covered at this stage, such as structured identifiers, device metadata, behavioral logs, and user profile fields. In production, Tier-1 flags roughly 60--70\% of private blocks while adding negligible compute overhead.

\paragraph{\textbf{Tier 2: General Privacy Detector}}
Tier-1 is fast but incomplete: pattern rules cannot reliably capture privacy signals that are implicit (e.g., ``my email is Kevin, dot, James, at Gmail, dot com—please don’t share it''), paraphrased, or expressed in diverse formats and languages. Tier-2 extends coverage with a lightweight semantic classifier that runs at low latency and high throughput.

Our first attempt was to use models that are explicitly fine-tuned for PII detection, since they are designed for this task and are easy to deploy. We benchmarked several fine-tuned transformer detectors for accuracy, multilingual coverage, and inference cost (\autoref{tab:pii-model-comparison}). Models such as \textit{DistilBERT-PII} and \textit{PII-BERT-base} exceed 93\% accuracy, and \textit{Piiranha-v1} reports 99.4\% accuracy with 125M parameters. However, these detectors are often trained for token-level labeling, which leads to fragmented outputs and makes them brittle when the input format changes. In our tests, their generalization to unseen categories is weak: on a \texttt{pii-masking} subset with 16 novel PII types, \textit{Piiranha-v1} detects only 33.4\%. This gap is problematic for multi-tenant serving, where tenants can introduce new formats and previously unseen sensitive fields.

\begin{figure*}
    \centering
    \includegraphics[width=0.75\linewidth]{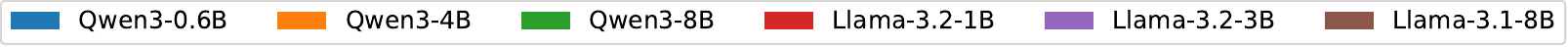}\\[-1.0em]
    \subfloat[Accuracy]{
        \includegraphics[width=0.24\linewidth]{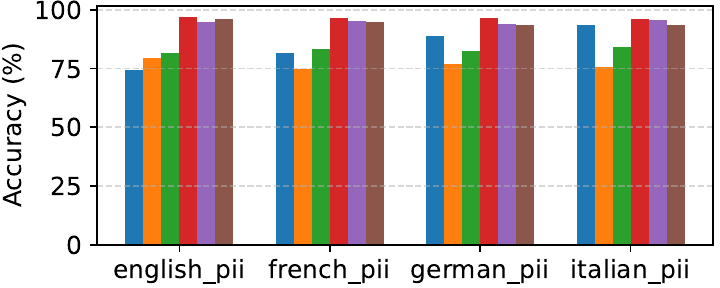}
    }
    \subfloat[Mean Latency]{
        \includegraphics[width=0.24\linewidth]{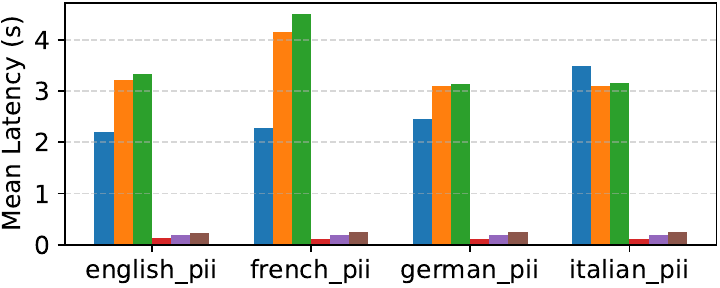}
    }
    \subfloat[P95 Latency]{
        \includegraphics[width=0.24\linewidth]{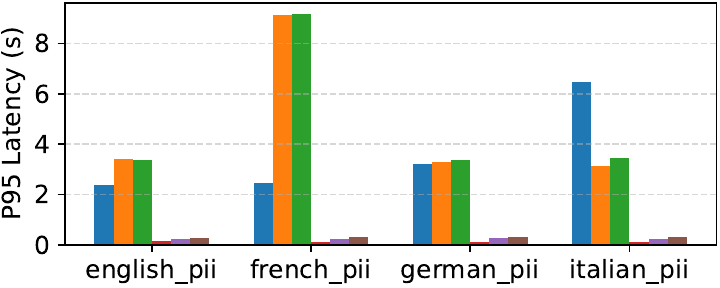}
    }
    \subfloat[P99 Latency]{
        \includegraphics[width=0.24\linewidth]{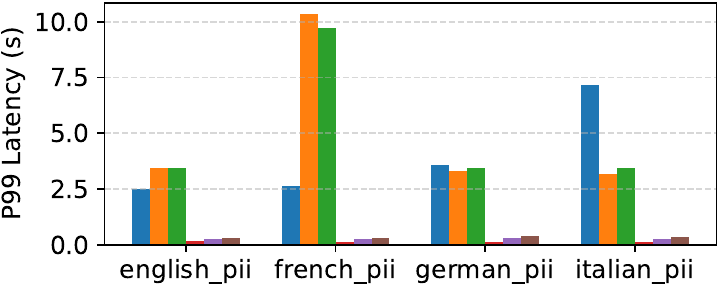}
    }
    \caption{Latency of lightweight general LLM models for PII detection}
\label{fig:level2_latency}
\vspace{-1.0em}
\end{figure*}

We then moved to lightweight LLMs, since the hard cases in Tier-2 are driven by semantics rather than surface patterns. We evaluated compact models from the Qwen3 and Llama-3 families for accuracy, out-of-distribution tolerance, and latency. As shown in~\autoref{fig:level2_latency}, \textbf{Llama-3.2-1B} offers the best accuracy--latency trade-off in our setting: it reaches near-perfect detection accuracy while keeping inference latency close to smaller transformer detectors. It also maintains stable performance across six languages. Based on these results, \systemname\ uses Llama-3.2-1B as the Tier-2 detector in the following experiment.

\begin{table}[t]
\centering
\caption{Comparison of lightweight fine‑tuned models for PII detection}
\vspace{-0.5em}
\resizebox{\columnwidth}{!}{%
\begin{tabular}{lccccc}
\toprule
\textbf{Model} & \textbf{Base Arch.} & \textbf{Size} & \textbf{Accuracy} & \textbf{Token vs Seq} & \textbf{Langs / Types} \\
\midrule
DistilBERT‑PII~\cite{distillbert-pii} & DistilBERT‑base & 66M & 95.22\% & Token-level & 1 lang / 5 types \\
Piiranha‑v1~\cite{piiranha-v1} & DeBERTa‑v2‑base & 125M & 99.44\% & Token-level & 6 langs / 17 types \\
PII‑BERT‑base~\cite{pii_model} & BERT‑base‑cased & 110M & 99.11\% & Token-level & English / gen data \\
dbert‑pii‑det.~\cite{dbert_pii_det} & DistilBERT‑uncased & 66M & 94.33\% & Token-level & mixed syn. \\
DePrompt~\cite{sun2024deprompt} & ChatGLM2‑6B & 6B & 95.95\% & Sequence-level & Chinese \\
GPT‑4o‑mini~\cite{shen2025enhancing} & GPT‑4o‑mini & 1B & 98.95\% & Sequence-level & English / Edu \\
% LLM‑Anonymizer & Llama‑3‑70B & 70B & 98.05\% & Sequence-level & Medical letters \\
\bottomrule
\end{tabular}%
}
\label{tab:pii-model-comparison}
% \vspace{-1.0em}
\end{table}

\paragraph{\textbf{Tier 3: Context-Aware Validation}}
Tier-1 and Tier-2 operate mainly on local text spans, so they can fail when privacy sensitivity depends on earlier turns. This happens when the current text is not identifying by itself, but becomes identifying once combined with prior dialogue (e.g., a name or location revealed earlier). Tier-3 targets these context-dependent cases and re-checks the block with an explicit view of recent conversation history.

Tier-3 is triggered only for \emph{Tier-2 low-confidence} decisions. For each such block, \systemname\ constructs a validation prompt that includes the current input and a bounded window of recent dialogue history, and then asks the \emph{in-service} LLM to decide whether the block should remain \texttt{Private} or can be promoted to \texttt{Shareable}. We use the serving LLM for this stage because the decision often requires long-context semantic reasoning; deploying a separate validation model would add extra replicas and deployment cost. Since Tier-3 runs only on Tier-2 low-confidence cases, it is rarely on the hot path: in our experiments, fewer than 8\% of requests reach Tier-3.

\subsubsection{Fallback Protection and Attack Mitigation}
\label{sec:det-fallback}

Despite employing a hybrid, multi-stage detection pipeline, \systemname\ acknowledges that privacy classification may occasionally fail, due to nuanced semantic ambiguity, incomplete pattern coverage, or the model's limitations. To mitigate these residual risks after deployment, \systemname\ introduces \textbf{Safeguard}, a lightweight runtime fallback mechanism that continuously monitors KV-cache access behavior and promptly reacts to suspicious reuse patterns.

At the core of this runtime defense is a lightweight statistical monitor that continuously observes access behaviors to each KV-cache block. Specifically, \systemname\ maintains a rolling window of metadata for each entry, recording the current hit count ($hit\_cur$), the number of unique user identifiers ($u\_cnt$), the previous hit count (\textit{hit\_pre}), and the previous number of unique users (\textit{u\_pre}). These values are used to compute the \textbf{Reuse Diversity Ratio (RDR)}, \( \mathrm{RDR} = u\_cnt / hit\_cur \), which summarizes access dispersion: low RDR indicates concentrated access (e.g., frequent hits by a single user), while high RDR suggests broad usage across accounts.

When Safeguard observes a significant shift in reuse behavior, it applies a history-aware policy. If the block has historically exhibited broad reuse (large $u_{\mathrm{pre}}$), an increase in dispersion is typically benign and the block remains shareable. In contrast, if historical reuse is minimal (e.g., $u_{\mathrm{pre}}\approx 1$) yet the current window exhibits elevated dispersion (i.e., RDR exceeds a configurable threshold derived from our RDR-based policy), Safeguard treats the pattern as suspicious and \emph{immediately demotes} the block from shareable to private. This demotion prevents further cross-tenant reuse and removes the exploitable timing signal induced by cache hits.

\subsubsection{Asynchronous Detection and Streaming-Aware Scheduling}
\label{sec:det-async}

To keep privacy detection from increasing inference latency, \systemname\ decouples classification from the critical execution path of LLM serving. When a KV-cache block is created, the system immediately inserts it as \texttt{Private} and continues serving the request without waiting for any detector output. Classification runs asynchronously in worker threads that pull pending blocks from a queue, batch them, and execute Tier-1/2/3 detection on GPUs when applicable. Batching amortizes per-call overhead and lets the detector use GPU parallelism without stalling the serving threads.

\systemname\ adapts its promotion policy to system conditions using a load-aware threshold. Under heavy traffic or when the runtime monitor flags probe-like access patterns, \systemname\ tightens the decision boundary and promotes fewer blocks, which favors conservative isolation. Under normal load, it relaxes the boundary to increase the shareable portion of the cache and recover more reuse benefit. Once a block is verified as \texttt{Shareable}, \systemname\ atomically flips its visibility bit and updates the corresponding prefix-tree metadata, so subsequent lookups can reuse the block across tenants without inconsistent intermediate states.

% This design keeps TTFT stable by removing synchronous classification from request handling, while still allowing the cache to become more shareable over time as background verification completes.

\subsection{Privacy-Aware Cache Management}
\label{sec:system}

\begin{figure}[t]
    \centering
    \includegraphics[width=0.95\columnwidth]{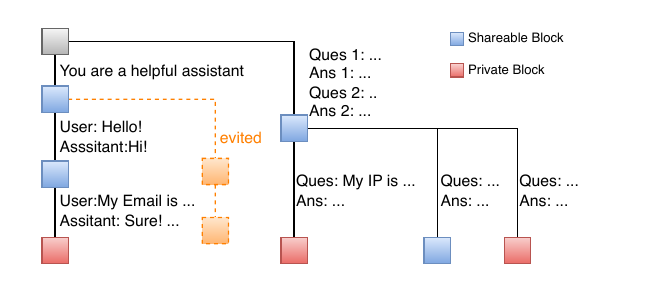}
    \caption{The Unified Privacy/Shareable Cache Index Tree.}
    \label{fig:unified-cache}
    \vspace{-1.0em}
\end{figure}

After introducing the privacy detection subsystem, \systemname\ can reliably determine whether a user’s input contains private information. The next question is how to use these privacy labels during KV-cache reuse: how to store, index, and evict blocks while privacy boundaries are enforced. We therefore focus on the memory subsystem in this subsection. The design centers on a unified radix-tree index with fine-grained access control, compression-aware private paths, and a progressive eviction mechanism that maintains memory efficiency under multi-tenant workloads.

\subsubsection{Unified Privacy-Preserving Cache Index.}
To distinguish private and shareable KV-cache blocks, a natural baseline is to maintain two separate caches. This split complicates memory accounting, duplicates common prefixes across the two structures, and makes eviction and tier migration harder to manage. To avoid this fragmentation, as shown in~\autoref{fig:unified-cache}, \systemname\ adopts a single radix-tree index to organize KV-cache entries across HBM/DRAM/SSD. This unified hierarchy maintains both scalability and privacy integrity. Each node stores two metadata fields: \texttt{private\_tag} (0 for shareable, 1 for private) and \texttt{creator\_id}. These fields define a visibility rule enforced during lookups: shareable nodes are accessible to all users, while private nodes are restricted to the creator. The rule is checked at each hop, so cross-tenant access is blocked even if private and shareable prefixes overlap in the tree.

\begin{itemize}[leftmargin=1.5em, itemsep=1pt, topsep=2pt]
\item \textbf{Insert.} \systemname\ appends newly generated KV-cache blocks along the radix-tree path. Each new node is created with $\texttt{private\_tag}$ =1 by default until it is verified by the detection pipeline, and it records the corresponding \texttt{creator\_id}.
\item \textbf{Search.} A lookup traverses the prefix path and checks the $\texttt{private\_tag}$ at each hop. If the node is private, the requester must match the stored $\texttt{creator\_id}$; otherwise, the traversal stops. This check prevents cross-tenant reuse even when private and shareable prefixes overlap.
\item \textbf{Evict.} Eviction follows an epoch-based LRU policy. Shareable nodes use reference counting to protect shared paths, while private nodes are pruned gradually to avoid removing entire tenant-specific branches too early.
\end{itemize}

\begin{figure}[t]
    \centering
    \includegraphics[width=0.9\columnwidth]{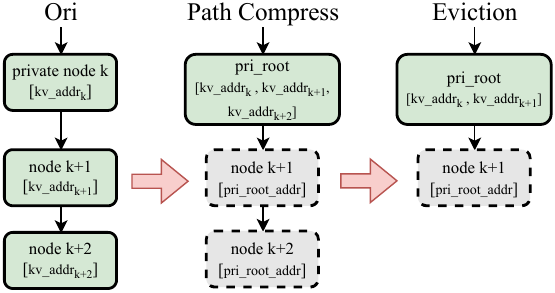}
    \caption{An Example of \textit{Path Compression} and \textit{Progressive Eviction}.}
    \label{fig:path-compression}
    \vspace{-1.0em}
\end{figure}

\subsubsection{Private Tree Optimization and Progressive Eviction}

In a unified radix-tree, shareable nodes often branch because many tenants reuse the same prefixes, while private nodes tend to form long, mostly linear paths tied to a single tenant. This asymmetry creates two practical issues: (i) private-heavy lookups can incur deep pointer chasing, and (ii) naive eviction may prioritize deleting entire large private branches. To address these issues, \systemname\ adds two mechanisms for private paths: \emph{path compression} to reduce traversal overhead and \emph{progressive eviction} to reclaim space without removing useful prefixes.

\paragraph{Path Compression}
\systemname\ SafeKV compresses consecutive private nodes belonging to the same tenant to reduce traversal depth on long private chains (~\autoref{fig:path-compression}). The system collapses a private chain into one composite node, denoted as \textit{pri\_root}. This node aggregates the KV addresses of the compressed descendants into a single vector and sets \textit{is\_compressed = true}. During lookup, traversal stops at \textit{pri\_root}; the system reads the required KV blocks directly from the address vector, which reduces pointer chasing and avoids deep traversal.

Compression does not remove the logical structure needed for lifecycle management. Compressed descendants remain in the metadata hierarchy but are marked inactive (\textit{after\_compress = true}) and linked back to \textit{pri\_root} via lightweight references. \textit{pri\_root} also records the aggregate memory footprint of its compressed subtree so the memory manager can charge usage and make eviction decisions without expanding the chain.

\paragraph{Progressive Eviction}
To avoid coarse-grained eviction of private branches, \systemname\ reclaims private subtrees incrementally from leaf to root. The system maintains a global \textit{epoch\_counter}, and each node stores its most recent access epoch. The evictor first prunes inactive leaves, which frees space while keeping the prefix context that may still be useful. For a compressed private chain, reclamation removes stale descendant entries from the \textit{pri\_root} address vector incrementally. The \textit{pri\_root} becomes eligible for eviction only after all subentries in its address vector have been released. This policy frees idle segments early without tearing down the entire tenant-local prefix.

% \paragraph{Epoch-Based LRU with Privacy-Aware Priority.}
% \systemname\ maintains a coarse-grained global epoch counter to approximate LRU ordering. Nodes with larger epoch deltas are evicted first. When multiple candidates have similar age, shareable nodes are evicted before private ones, since private caches typically have lower reuse probability but higher isolation requirements. This privacy-aware priority minimizes premature exposure while keeping memory usage predictable.

% These optimizations support efficient compression and controlled eviction of private KV-cache entries, aligning memory usage with the dynamic requirements of multi-tenant LLM inference systems.
\section{Implementation}
\label{sec:implement}

We implement \systemname\ as an extension to SGLang’s serving runtime; we add a privacy-aware KV-cache manager and a small set of control-plane services while leaving the model execution path and request API unchanged. Specifically, we extend SGLang’s radix-tree prefix cache with per-block privacy labels and tenant-ownership metadata, \systemname\ checks these tags on every lookup and reuse decision. We also add private-path compression and a progressive eviction policy to the radix-tree index to preserve memory efficiency under multi-tenant workloads.  The detection pipeline runs off the critical path in a dedicated manager: new KV-cache blocks are marked private on insertion, then processed in batches by lightweight detector (rules and a small classifier) and escalated to an LLM-based validator only for ambiguous cases. Once a block is validated, it is promoted to shareable without stalling decoding. \systemname\ remains compatible with SGLang’s existing caching and scheduling interfaces: it requires no changes to upstream request handlers or deployment tooling, and privacy isolation can be enabled or disabled via configuration.
\section{Security Analysis}
\label{sec:security}

\paragraph{Adversary Model and Security Goals}
\systemname\ targets timing-based side-channel leakage caused by cross-tenant KV-cache reuse in multi-tenant LLM serving, as described in~\autoref{sec:threat_model}. We consider a black-box, API-only adversary who uses repeated probes and statistical tests on TTFT to distinguish cache hits from misses. Our design goals are: (G1) block cross-tenant reuse of KV blocks that contain privacy-sensitive content, (G2) preserve reuse benefits for prefixes that are verified safe, and (G3) limit the impact of  misclassification using runtime safeguards.

\paragraph{Leakage Surface and Security Property}
The leakage surface we focus on is \emph{API-visible TTFT distinguishability} that arises from cross-tenant cache hits. \systemname\ enforces reuse using an explicit visibility state for each cached prefix. Let $r_v=\{t_1,\ldots,t_k\}$ be a victim prefix. Let $C(t_1,\ldots,t_k)$ indicate whether the corresponding KV block is \texttt{Shareable} to other tenants. A probing request $r_a$ with the same prefix can exhibit a cache-hit TTFT only when $C=\texttt{True}$:
\[
\text{TTFT}(r_a) \in
\begin{cases}
T_{\text{hit}} & \text{if } C(t_1,\ldots,t_k)=\texttt{True},\\
T_{\text{miss}} & \text{otherwise.}
\end{cases}
\]

\systemname\ inserts new blocks as \texttt{Private} by default and sets $C=\texttt{True}$ only after the block passes multi-tier detection and is explicitly promoted to \texttt{Shareable}. Consequently, sensitive prefixes eliminate cross-tenant hit signals unless they are incorrectly promoted.

\paragraph{Misclassification Risk and Exposure Control}
Let $S$ denote the event that a KV block is privacy-sensitive. The only way for cross-tenant reuse of a sensitive block is a false negative that leads to promotion. Define $\alpha_1$ as the false-negative rate of Tier-1. Define $\alpha_{2|1}$ as the false-negative rate of Tier-2 conditioned on passing Tier-1, and $\alpha_{3|12}$ as the false-negative rate of Tier-3 conditioned on passing Tier-1 and Tier-2. Then the probability that a sensitive block becomes shareable is
\[
P_{\text{leak}} \triangleq P(C=\texttt{True}\mid S)
= \alpha_1 \cdot \alpha_{2|1} \cdot \alpha_{3|12},
\]
where each term is measured in~\ref{sec:evaluation}.

Even when a sensitive block is mistakenly promoted, \systemname\ treats sharing as revocable and limits the exposure window using runtime monitoring. The safeguard tracks reuse dispersion per block using the reuse diversity ratio (RDR),
\[
\mathrm{RDR}_b = \frac{u_{\text{cnt}}}{\text{hit}_{\text{cnt}}},
\]
computed over a fixed window, where $u_{\text{cnt}}$ is the number of distinct tenants that accessed block $b$ in the window and $\text{hit}_{\text{cnt}}$ is the total number of hits. Probe-heavy access patterns typically increase $\text{hit}_{\text{cnt}}$ faster than $u_{\text{cnt}}$, which lowers $\mathrm{RDR}_b$. If $\mathrm{RDR}_b$ stays below a threshold for consecutive windows, the safeguard demotes the block (\texttt{Shareable}$\rightarrow$\texttt{Private}) and enqueues it for re-validation. After demotion, future cross-tenant hits on that block stop.

\paragraph{Limitations and Out-of-Scope Attacks}
\systemname\ reduces TTFT distinguishability that is attributable to cross-tenant reuse of sensitive prefixes. It does not provide cryptographic secrecy, and it does not address attackers with privileged access (e.g., kernel, hypervisor, or direct memory read). We also do not attempt to handle side channels outside the TTFT surface, such as network jitter, token-count leakage, or other co-resident system attacks. Our evaluation therefore focuses on measurable reductions in hit/miss distinguishability under repeated-query attackers that match prior prefix/KV-cache probing settings.

\section{Evaluation}
\label{sec:evaluation}

In this section, we evaluate \systemname\ on a diverse set of recent LLMs, including Phi-4~\cite{phi-4}, Qwen3-30B-A3B~\cite{qwen3-30b-a3b}, Qwen3-32B~\cite{qwen3-32b}, Qwen3-235B-A22B~\cite{qwen3-235b-a3b}, Llama-3.3-70B~\cite{llama70b} and DeepSeek-R1~\cite{deepseek-r1}. We focus on three research questions:
\begin{itemize}[leftmargin=1.5em, itemsep=1pt, topsep=2pt]
    \item \textbf{[RQ1] Effectiveness}: Does \systemname\ suppress TTFT-based distinguishability for \emph{sensitive} prompts under cross-tenant KV-cache reuse? (~\autoref{sec:rq1})
    \item \textbf{[RQ2] Cost}: What runtime overhead does \systemname\ introduce, including the latency and throughput impact of privacy detection? (~\autoref{sec:rq2})
    \item \textbf{[RQ3] Performance}: Compared to recent isolation baselines for KV-cache timing defenses, how much throughput/latency benefit does \systemname\ retain under realistic workloads? (~\autoref{sec:rq3})
\end{itemize}

\subsection{Experiment Setup.}

\noindent\textbf{Testbed.}
We evaluate the \systemname\ prototype described in~\autoref{sec:implement}. Unless stated otherwise, all experiments run on one server with $8\times$NVIDIA H20 GPUs (96\,GB each). We report the mean across repeated runs and include P95/P99 when latency tails matter.

\noindent\textbf{Workload.}
We use two groups of workloads. For privacy-sensitive inputs, we use the \texttt{pii-masking} workload with multilingual samples spanning 54 PII categories. This workload drives the detector pipeline and provides the sensitive-prompt cases used in the security evaluation. For performance comparisons (RQ3), we use three serving patterns that reflect common deployments. \emph{Single-Request PII} contains short privacy-sensitive queries. \emph{Multi-Turn Chat} contains conversational sessions with embedded PII and repeated prefixes within a session. \emph{System Prompt} contains requests that share a common system prompt across users and include user-specific PII in the remainder of the prompt.

\noindent\textbf{Security Evaluation Setup.}
We evaluate \systemname\ under the black-box, API-only adversary defined in~\autoref{sec:threat_model}. The attacker can create multiple accounts, issue requests repeatedly, and adapt future queries based on observed timing. The attacker observes TTFT and aims to infer whether a victim’s prefix has been cached through cross-tenant KV reuse. Our evaluation follows this end-to-end objective. In each experiment, we run attacker sessions for each target sensitive prefix drawn from the \texttt{pii-masking} workload. The attacker generates candidate queries without access to the victim’s plaintext. The attacker replays prefixes, collects TTFT traces under the live serving stack, and applies statistical analysis to decide whether the target prefix is cached. Finally, we report the defense success rate and ROC-AUC across models to quantify access success.

\noindent\textbf{Performance Evaluation Setup.}
We compare \systemname\ against three representative KV-cache management strategies: 
\begin{itemize}[leftmargin=1.5em, itemsep=1pt, topsep=2pt]
    \item \textbf{(1). SGLang~\cite{sglangcode}:} Full global cache sharing with no privacy protection, achieving maximum efficiency; 
    \item \textbf{(2). Cache-Partition~\cite{pang2024cache}:} A state-of-the-art defense for KV-cache timing side channel attacks that isolates KV-cache per user via separate radix trees, eliminating cross-user reuse to achieve strong privacy; 
    \item \textbf{(3). Shared System Prompt:} A conservative and widely adopted practice that shares KV-cache only for a fixed system prompt across users, while keeping user-provided prompts isolated to avoid cross-tenant exposure. 
\end{itemize}
We report throughput and TTFT on the three performance workloads above.

\subsection{\textbf{Security Against Timing Probing}}
\label{sec:rq1}
This section evaluates whether \systemname\ suppresses API-visible TTFT timing signals that enable cross-tenant cache probing on sensitive prefixes.

\begin{figure}
    \centering
    \includegraphics[width=0.95\linewidth]{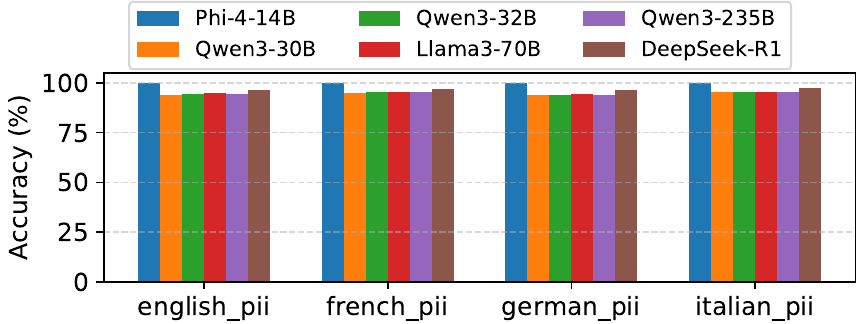}
    \caption{Defense success rate of \systemname\ under Timing Side-Channel Attacks.}
\label{fig:defense_acc}
\vspace{-0.5em}
\end{figure}

\begin{figure}
    \centering
    \includegraphics[width=0.95\linewidth]{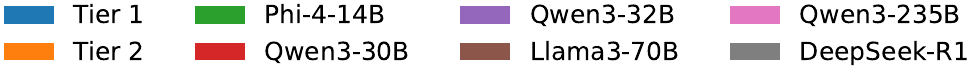}\\[-0.01em]
    \subfloat[Accuracy]{
        \includegraphics[width=0.95\linewidth]{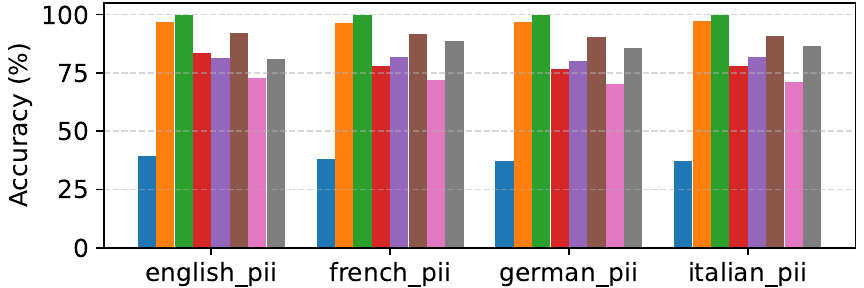}
    }
    \hspace{0.01\textwidth}
    \subfloat[Accuracy with Complex Prompts]{
        \includegraphics[width=0.95\linewidth]{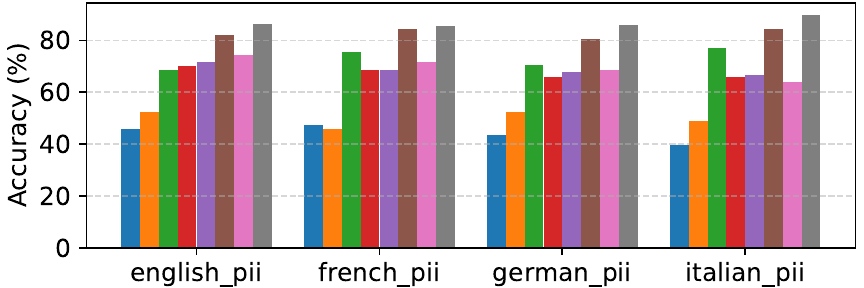}
    }
    \caption{Comparison of the Accuracy of Multi-Tier Privacy Detectors under simple/complex requests}
\label{fig:level3_accuracy}
\vspace{-1.0em}
\end{figure}

\subsubsection{\textbf{Overall Defense Effectiveness.}}
We first evaluate \systemname's ability to mitigate timing-based KV-cache side-channel attacks. \autoref{fig:defense_acc} summarizes the results across multiple model backbones. Overall, \systemname\ achieves an average defense success rate exceeding 94\% via its three-tier privacy pipeline, which identifies and suppresses the majority of malicious probes during inference. Here, defense success is defined as the fraction of probe windows in which the attacker cannot reliably distinguish sensitive-hit from sensitive-miss TTFT. Moreover, the effectiveness of \systemname\ improves as the model and the service mature over time. For example, on a privacy-sensitive dataset spanning four languages, DeepSeek-R1 attains a defense success rate of 96.24\%-96.90\%; Microsoft Phi-4 is nearly fully robust to adversarial probing, consistent with its security-oriented post-training. These results indicate that \systemname\ provides an adaptive defense mechanism that remains effective under evolving workloads and model deployments.

Importantly, in \systemname, static classification errors are not equivalent to cross-tenant exposure. Each newly created KV block is inserted as \emph{Private} by default and is therefore ineligible for cross-tenant reuse; a block becomes shareable only after the asynchronous pipeline explicitly promotes it to \emph{Shareable}. Promotion is also revocable: \systemname's runtime feedback loop continuously monitors cross-tenant access patterns and demotes (re-privatizes) KV blocks that exhibit suspicious reuse. As a result, static classification primarily affects how quickly reuse becomes available, whereas the promotion gate and revocation determine whether cross-tenant reuse can persist under attack. Under sustained probing, the adaptive demotion mechanism prevents previously attacked entries from remaining shareable, driving the effective protection rate toward near-complete coverage.

\begin{figure}
    \centering
    \includegraphics[width=0.9\linewidth]{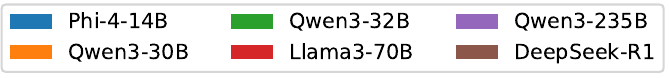}\\[-0.05em]
    \subfloat[Mean Latency]{
        \includegraphics[width=0.9\linewidth]{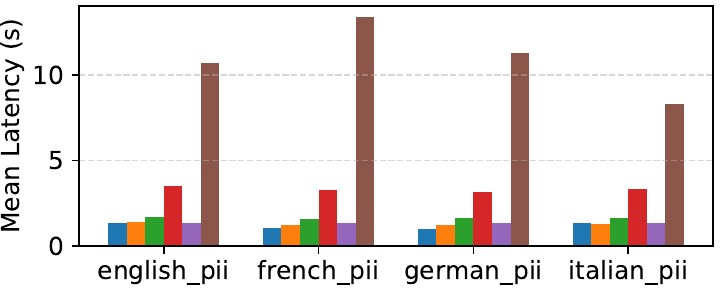}
    }
    \hspace{0.1\textwidth}
    \subfloat[P95 Latency]{
        \includegraphics[width=0.9\linewidth]{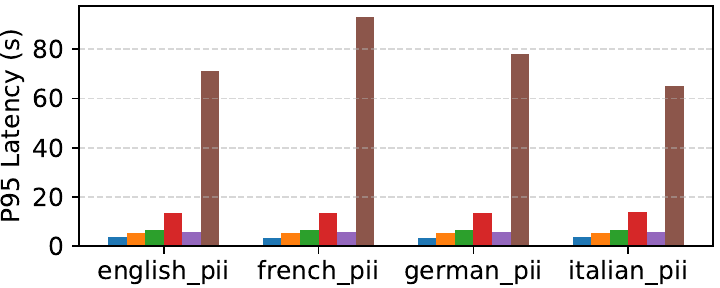}
    }
    \hspace{0.1\textwidth}
    \subfloat[P99 Latency]{
        \includegraphics[width=0.9\linewidth]{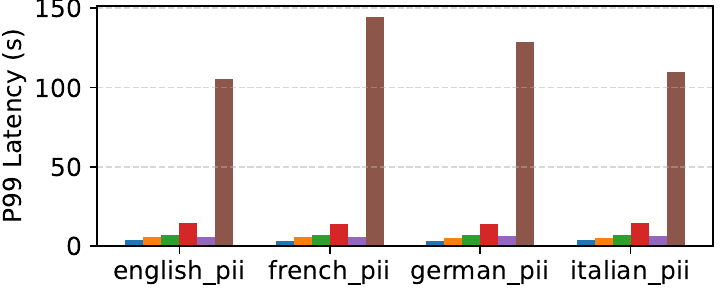}
    }
    \caption{Latency of PII detection for large-scale LLMs}
\label{fig:defense_latency}
\vspace{-0.5em}
\end{figure}

\subsubsection{\textbf{Accuracy of Multi-Tier Detection.}}
% We now evaluate the accuracy and effectiveness of \systemname’s hybrid privacy detection pipeline, which operates in three stages: Tier-1 rule-based matching, Tier-2 lightweight general privacy detector, and Tier-3 context-aware validation.
The hybrid pipeline complements fast pattern checks with semantic detection and context validation:

\textbf{Tier-1: Rule-Based Pattern Matching.}
As outlined in Section~\ref{sec:det-tiered}, the first stage performs fast keyword and regular expression matching to capture structured and explicit sensitive content. On multilingual test sets, the Tier-1 engine achieves detection accuracies of 39.24\%, 38.09\%, 37.35\%, and 37.33\% respectively (as shown in~\autoref{fig:level3_accuracy}(a)). While lightweight, this tier lacks semantic understanding and demonstrates limited accuracy.

\textbf{Tier-2: General Privacy Detector.}
Tier-2 employs a compact transformer-based language model to detect privacy-sensitive content that escapes pattern matching. Among candidates (\autoref{fig:level2_latency}), we choose \textbf{Llama-3.2-1B} as the default Tier-2 detector to achieve a balance between detection accuracy, inference latency, and GPU resource utilization. It achieves 96.85--97.15\% across languages while maintaining low latency.

\textbf{Tier-3: Context-Aware Validation.}
Tier-3 invokes the in-service LLM to resolve hard, context-dependent cases. As shown in \autoref{fig:level3_accuracy}, on regular PII-style prompts the lightweight Tier-2 detector can be more accurate than most of the in-service model; however, when longer prompts include richer context, Tier-2 accuracy drops noticeably. In contrast, stronger in-service models such as DeepSeek-R1 remain robust and can lift accuracy from roughly 50\% (Tier-2) to above 90\%. This gap highlights why Tier-3 is necessary for subtle privacy risks that only become apparent given broader context.

\begin{figure}
    \centering
    \includegraphics[width=0.9\linewidth]{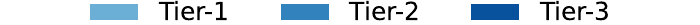}\\[-0.7em]
    \subfloat[Phi-4-14B]{
        \includegraphics[width=0.32\linewidth]{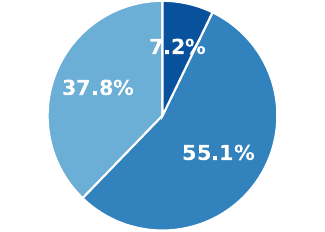}
    }
    \subfloat[Qwen3-30B]{
        \includegraphics[width=0.32\linewidth]{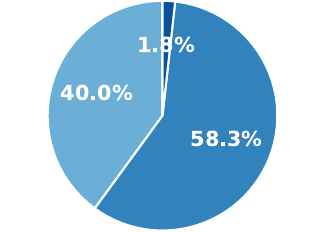}
    }
    \subfloat[Qwen3-32B]{
        \includegraphics[width=0.32\linewidth]{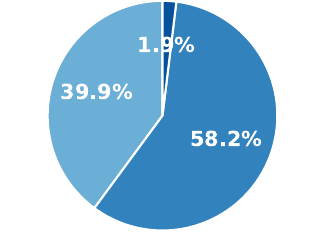}
    }
    \hspace{0.1\textwidth}
    \subfloat[Llama-3.3-70B]{
        \includegraphics[width=0.32\linewidth]{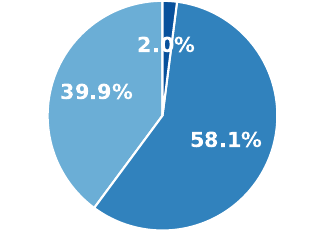}
    }
    \subfloat[Qwen3-235B]{
        \includegraphics[width=0.32\linewidth]{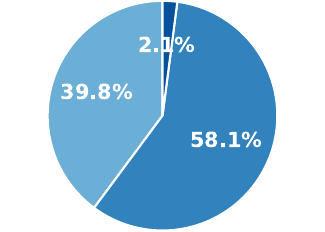}
    }
    \subfloat[DeepSeek-R1]{
        \includegraphics[width=0.32\linewidth]{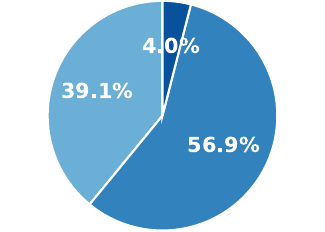}
    }
    \caption{Workload ratio of multi-tier detectors across different large-scale LLMs.}
\label{fig:defense_partition}
\vspace{-1.0em}
\end{figure}

\subsubsection{\textbf{Safeguard Robustness and Fallback Protection.}}
\begin{figure}
    \centering
    \subfloat[Phi-4-14B]{
        \includegraphics[width=0.46\linewidth]{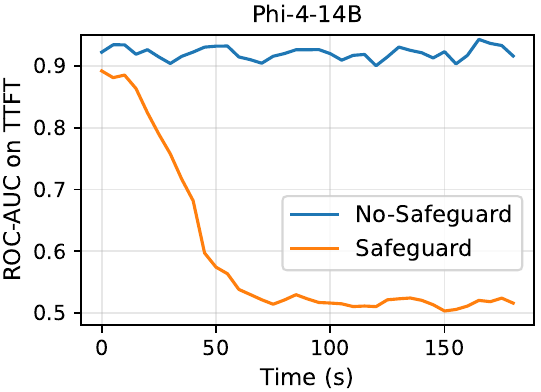}
    }
    \subfloat[Qwen3-30B]{
        \includegraphics[width=0.46\linewidth]{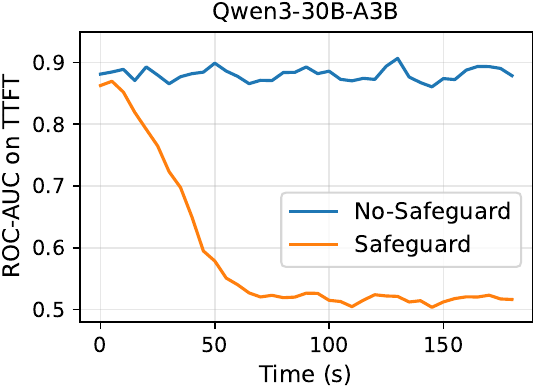}
    }
    \hspace{0.1\textwidth}
    \subfloat[Qwen3-32B]{
        \includegraphics[width=0.46\linewidth]{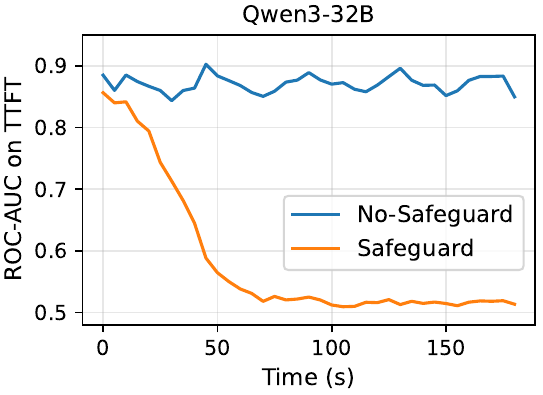}
    }
    \subfloat[LLama-3.3-70B]{
        \includegraphics[width=0.46\linewidth]{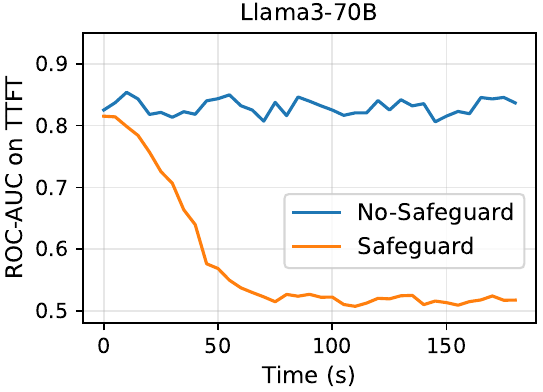}
    }
    \caption{Attack distinguishability over time measured as ROC-AUC on TTFT using a time window ($W=5s$). Without Safeguard, cache-hit vs.\ cache-miss TTFT remains separable, yielding consistently high AUC. With Safeguard enabled, once the RDR drops the configured threshold, the corresponding KV blocks are demoted (shareable$\rightarrow$private), collapsing AUC to $\approx 0.5$ and keeping it stable despite continued probing.
}
\label{fig:safeguard_auc_time}
% \vspace{-1.0em}
\end{figure}

False negatives in multi-tier privacy detection can lead to incorrect promotion of a sensitive KV-cache block and re-enable cross-tenant reuse. We therefore evaluate whether \systemname's Safeguard can terminate TTFT-based probing under the black-box, API-only adversary model in~\autoref{sec:threat_model}. In this experiment, the attacker repeatedly issues queries for a target prefix and uses TTFT to statistically distinguish cache hits from misses. Safeguard operates on a sliding window of length $W$ and demotes a KV block from \texttt{Shareable} to \texttt{Private} once the RDR-based trigger condition is satisfied. Figure~\ref{fig:safeguard_auc_time} reports ROC-AUC computed from TTFT samples within each window for probing-style traces. Without Safeguard, AUC remains high over time, indicating a persistent hit--miss timing signal. With Safeguard enabled, AUC drops to $\approx 0.5$ after the trigger and stays near random guess thereafter, indicating that Safeguard stops the exploitable signal rather than only weakening it.

\begin{table*}[ht]
\centering
% \small % 字体缩小
\setlength{\tabcolsep}{6.5pt} % 调整列间距
\caption{Latency breakdown of Tier-1 and Tier-2 across four languages.}
\vspace{-0.5em}
\begin{tabularx}{\textwidth}{l|ccc|ccc|ccc|ccc}
\toprule
& \multicolumn{3}{c|}{\textbf{english\_pii}} 
& \multicolumn{3}{c|}{\textbf{french\_pii}} 
& \multicolumn{3}{c|}{\textbf{german\_pii}} 
& \multicolumn{3}{c}{\textbf{italian\_pii}} \\
\textbf{Tier} & Mean(ms) & P95(ms) & P99(ms) & Mean & P95 & P99 & Mean & P95 & P99 & Mean & P95 & P99 \\
\midrule
Tier-1 & 0.10 & 0.16 & 0.20 & 0.11 & 0.17 & 0.20 & 0.11 & 0.17 & 0.21 & 0.11 & 0.17 & 0.20 \\
Tier-2 & 124.93 & 152.14 & 171.85 & 113.03 & 125.68 & 127.17 & 113.91 & 126.31 & 129.54 & 116.12 & 128.77 & 140.77 \\
\bottomrule
\end{tabularx}
\label{tab:latency_1_2}
\vspace{-0.5em}
\end{table*}

\begin{figure}
    \centering
    \includegraphics[width=0.85\linewidth]{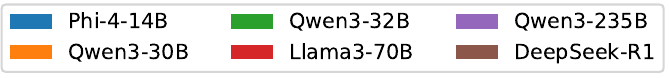}\\[-0.01em]
    \subfloat[Mean Latency]{
        \includegraphics[width=0.85\linewidth]{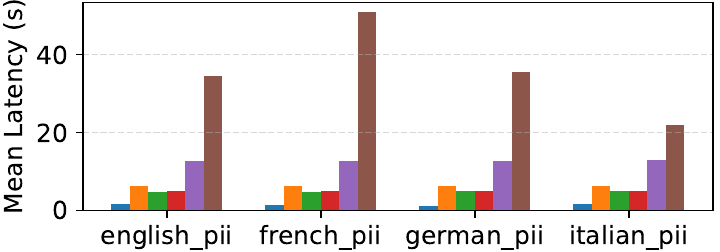}
    }
    \hspace{0.1\textwidth}
    \subfloat[P95 Latency]{
        \includegraphics[width=0.85\linewidth]{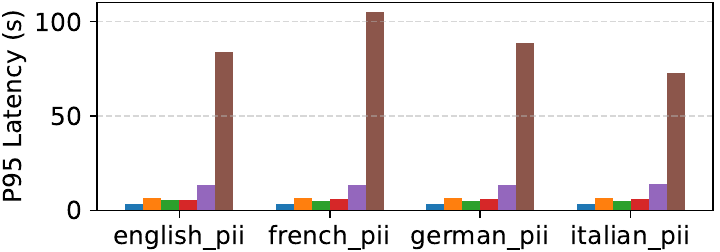}
    }
    \hspace{0.1\textwidth}
    \subfloat[P99 Latency]{
        \includegraphics[width=0.85\linewidth]{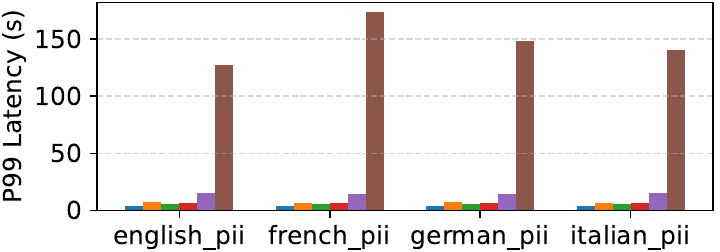}
    }
    \caption{The Latency breakdown of Tier-3 PII detection across four languages.}
\label{fig:level3_latency}
\vspace{-1.0em}
\end{figure}

\begin{figure}[t]
    \centering
    \includegraphics[width=0.9\linewidth]{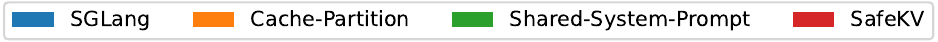}\\[-0.01em]
    \subfloat[Single-Request PII Mean Latency]{
        \includegraphics[width=0.9\linewidth]{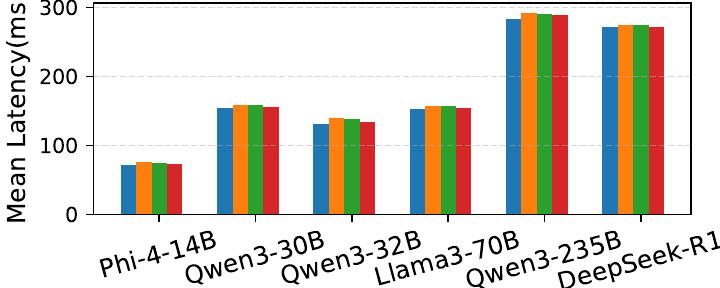}
    }
    \hspace{0.01\textwidth}
    \subfloat[Multi-Turn Mean Latency]{
        \includegraphics[width=0.9\linewidth]{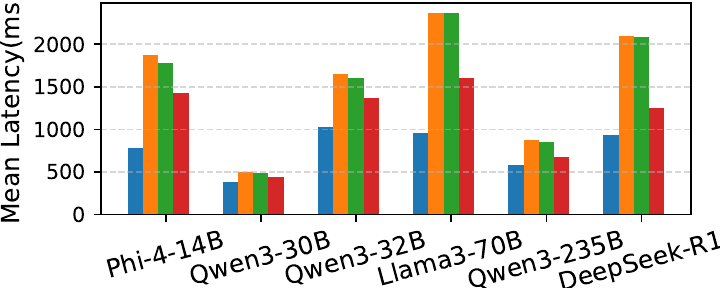}
    }
    \hspace{0.01\textwidth}
    \subfloat[System Prompt Mean Latency]{
        \includegraphics[width=0.9\linewidth]{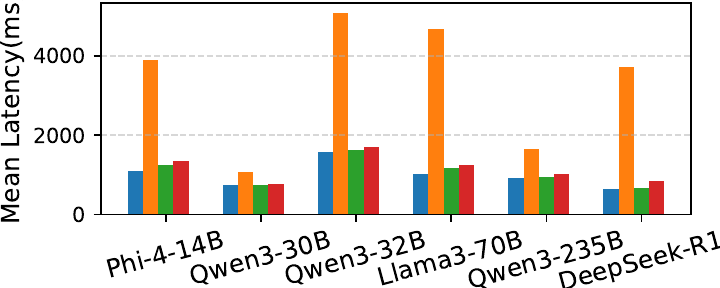}
    }
    % \hspace{0.1\textwidth}
    % \subfloat[Single Request PII P95 Latency]{
    %     \includegraphics[width=0.32\linewidth]{figures/pii_p95_latency.pdf}
    % }
    % \subfloat[Multi-Turn Chat P95 Latency]{
    %     \includegraphics[width=0.32\linewidth]{figures/multiturn_p95_latency.pdf}
    % }
    % \subfloat[System Prompt P95 Latency]{
    %     \includegraphics[width=0.32\linewidth]{figures/systemp_p95_latency.pdf}
    % }
    % \hspace{0.1\textwidth}
    % \subfloat[Single Request PII P99 Latency]{
    %     \includegraphics[width=0.32\linewidth]{figures/pii_p99_latency.pdf}
    % }
    % \subfloat[Multi-Turn Chat P99 Latency]{
    %     \includegraphics[width=0.32\linewidth]{figures/multiturn_p99_latency.pdf}
    % }
    % \subfloat[System Prompt P99 Latency]{
    %     \includegraphics[width=0.32\linewidth]{figures/systemp_p99_latency.pdf}
    % }
    \caption{Comparison of TTFT of \systemname\ with SGLang, Cache-Partition, and Shared-System-Prompt in different working scenarios}
\label{fig:performance}
\vspace{-1.0em}
\end{figure}

\subsection{\textbf{Overheads of Asynchronous Detection}}
\label{sec:rq2}

\paragraph{\textbf{Overhead of Multi-Tier Detection}}
We measure the end-to-end runtime cost introduced by the complete multi-tier privacy detection pipeline across several LLM backends. \autoref{fig:defense_latency} summarizes the average, P95, and P99 service latencies for the three-tier detection process. Although larger models incur higher analysis costs, this increase is primarily due to the Tier-3 inference stage. As shown in \autoref{tab:latency_1_2}, the rule-based detection in Tier-1 completes within 0.2 ms per prompt, while Tier-2, which employs the compact Llama-3.2-1B model, adds an average latency below 125 ms. In contrast, \autoref{fig:level3_latency} shows that Tier-3 exhibits substantially higher latency.

However, its overall impact on service time remains limited. As illustrated in \autoref{fig:defense_partition}, over 92\% of requests are fully classified by the first two tiers, leaving only a small fraction to be processed by the more expensive Tier-3 analysis. Furthermore, \systemname\ employs an asynchronous detection pipeline that decouples privacy classification from the token generation path. Consequently, privacy checks never block inference. Even when Tier-3 experiences long-tail latency, the effect is confined to delaying the promotion of certain KV-cache entries from private to shared state. This delay neither weakens ongoing privacy protection nor disrupts intra-session reuse; it only postpones cross-user sharing for a small subset of cache blocks. Given that Tier-3 handles fewer than 8\% of all requests, such worst-case overhead is well within acceptable operational bounds.

% \begin{figure*}
%     \centering
%     \subfloat[Mean Latency]{
%         \includegraphics[width=0.33\linewidth]{figures/avg_lat_vs_prompt_length.pdf}
%     }
%     % \hspace{0.1\textwidth}
%     \subfloat[P95 Latency]{
%         \includegraphics[width=0.33\linewidth]{figures/p95_lat_vs_prompt_length.pdf}
%     }
%     % \hspace{0.1\textwidth}
%     \subfloat[P99 Latency]{
%         \includegraphics[width=0.33\linewidth]{figures/p99_lat_vs_prompt_length.pdf}
%     }
%     \caption{The Latency of PII detection vs different prompt length}
% \label{fig:lat_vs_prompt_length}
% \vspace{-0.5em}
% \end{figure*}

\subsection{\textbf{Performance Under Multi-Tenant Workloads}}
\label{sec:rq3}
This section compares \systemname\ against isolation-based cache management and conservative sharing baselines under representative multi-tenant workloads. We report throughput and TTFT to show how much performance \systemname\ recovers while enforcing privacy-aware reuse.

\paragraph{\textbf{LLM Inference Latency}}
Our primary evaluation metric is \textit{inference latency}, particularly focusing on \textit{TTFT}, since TTFT effectively reflects prefill savings from KV-cache reuse. As depicted in~\autoref{fig:performance},  In single-request PII, reuse is scarce; all methods are similar. In multi-turn conversational scenarios (constructed using SharedGPT~\cite{sharegpt-chat} dialogues augmented with pii-masking~\cite{pii-masking} queries to embed privacy-sensitive information), substantial opportunities for prefix reuse emerge, resulting in notable performance variations. Against full isolation, \systemname\ cuts the extra TTFT overhead from 50.41\% to 11.74\% on Qwen-235B-A22B, and from 118\% to 34.28\% on DeepSeek-R1. Finally, in the system prompt scenario (pii-masking requests with a uniform system prompt (approximately 8192 tokens)), \systemname\ continues to perform effectively, although slightly behind the Shared-System-Prompt method explicitly optimized for this scenario. Overall, these results demonstrate that \systemname\ consistently achieves superior latency performance compared to Cache-Partition and remains adaptable across diverse real-world use cases.

\begin{figure}
    \centering
    \includegraphics[width=0.99\linewidth]{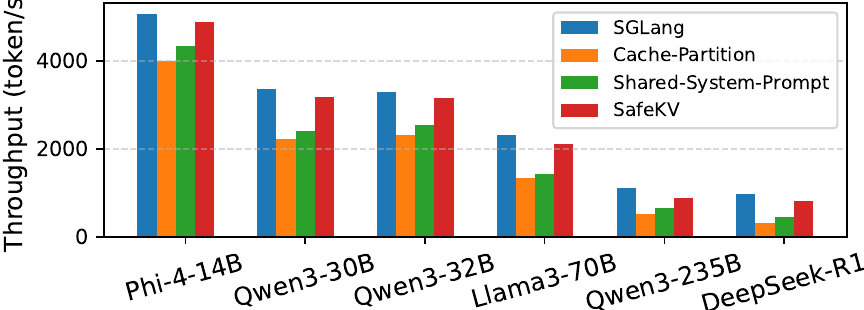}
    \caption{Comparison of Throughput of \systemname\ with SGLang, Cache-Partition, and Shared-System-Prompt}
\label{fig:throught}
\vspace{-1.5em}
\end{figure}

\paragraph{\textbf{LLM Inference Throughput}}
Throughput is critical for evaluating a system's capability to handle sustained high-volume requests. To measure this, we conduct throughput benchmarking at a fixed input rate of 16 RPS (request per second), reporting token throughput as the total number of tokens divided by the end-to-end inference time. Test workloads are derived from: \texttt{pii\_masking~\cite{pii-masking}}, \texttt{ShareGPT~\cite{sharegpt-chat}}, and the requests with uniform system prompt. As shown in~\autoref{fig:throught}, \systemname\ improves over Cache-Partition by $1.36\times$–$2.66\times$ across the evaluated workloads by isolating only sensitive paths and preserving reuse for shareable prefixes. Gains are larger on heavier foundation models (e.g., DeepSeek-R1), where LLM prefill dominates utilization.
\section{Discussion}
\label{sec:discussion}

% discussion 要充分一点
% \noindent\textbf{Protection Scope and Assumptions}
% \systemname\ targets the timing side-channel surface created by cross-tenant KV reuse, where cache hits and misses induce an API-visible TTFT gap (\autoref{sec:threat_model}). Our guarantee is therefore scoped to two outcomes: (i) privacy-sensitive KV blocks are not reused across tenants, and (ii) timing probes on sensitive prefixes become less reliable under the black-box adversary in~\autoref{sec:threat_model}. We do not address non-timing channels such as co-resident contention or microarchitectural leakage, which require isolation beyond the serving stack. We also assume the service’s tokenizer (and any normalization) defines the prefix identity used for reuse; when semantically similar inputs tokenize differently (e.g., Unicode variants or whitespace perturbations), reuse rates drop and detector behavior can vary across formats.

\noindent\textbf{Detection Robustness and Safeguard}
\systemname\ uses a three-tier detector to decide whether a block may become shareable (\autoref{sec:detection}, \autoref{sec:det-tiered}), and false negatives remain possible as privacy formats evolve. The design limits the consequence of such errors by construction: blocks are private by default, promotion is asynchronous, and promotion is revocable. The remaining risk is an incorrect promotion that enables cross-tenant reuse for a limited time window. Safeguard reduces this window by monitoring access dispersion and demoting blocks whose reuse pattern is consistent with probing (\autoref{sec:det-fallback}). In deployment, the demotion threshold should be tuned to the expected workload: aggressive demotion reduces exposure time but can also reduce reuse on benign popular prefixes.

\noindent\textbf{Deployment and Integration Considerations}
The main operational costs come from detector and the cache metadata required to enforce visibility and support demotion. Since detection is off the critical inference path (\autoref{sec:detection}), higher-tier slowdowns primarily delay private-to-shareable promotion rather than stalling generation. On the serving side, integration requires storing a per-block visibility state and creator binding, plus lightweight accounting for monitoring and demotion (\autoref{sec:system}). As a result, \systemname\ is easiest to adopt in stacks that already maintain structured KV indices (e.g., radix-tree or block-based caching), where these fields can be added without changing the request execution model.

\noindent\textbf{Generality and Transferability}
\systemname\ separates enforcement from classification: the cache indexing and visibility checks are model-agnostic, while the detector depends on the workload and privacy policy (\autoref{sec:system}, \autoref{sec:detection}). This separation makes the cache-management layer transferable across models and serving stacks that support prefix-indexed reuse. For multimodal systems, the enforcement layer carries over directly, but detection must expand to modality-specific signals (e.g., text extracted from images or image-derived identifiers). This suggests treating detection as a pluggable policy module while keeping cache enforcement unchanged.

\section{Related Work}
\label{sec:related_work}

\subsection{\textbf{Multi-tenant Security}}
Side-channel attacks have long threatened multi-tenant systems due to shared resource usage. They can be broadly categorized into classical system-level attacks and emerging model-level side channels. 

Classical side-channel attacks exploit shared hardware or OS abstractions in multi-tenant systems. Cross-VM attacks recover sensitive data between co-located virtual machines by monitoring CPU caches or memory access patterns~\cite{zhang2012cross,zhang2014cross,varadarajan2015placement,xiao2016one}. 
Other works leverage shared OS resources, such as OS data structure or public file systems, to launch cross-application attacks in Unix, Android, or iOS environments~\cite{zhang2009peeping,jana2012memento,chen2014peeking,zhang2016return,zhang2018level,wang2023danger}. Recent studies further reveal that multiple WebAssembly modules isolated in the same runtime are vulnerable to cross-module attacks~\cite{narayan2021swivel,kocher2019spectre}.

Complementing these, recent work highlights a new class of attacks targeting LLM inference backends. These attacks exploit timing differences caused by shared KV-cache reuse to infer user inputs~\cite{wu2025prompt,zheng2024inputsnatch,song2024early}. For example, PromptPeek~\cite{wu2025prompt} and InputSnatch~\cite{zheng2024inputsnatch} reconstruct user prompts via TTFT measurements in black-box settings. Unlike classical channels, these attacks are unique to LLM-serving pipelines, where performance optimizations inadvertently expose privacy risks. Our work builds on these findings by proposing a practical, multi-tier defense system that addresses this emerging attack surface.

\subsection{\textbf{Defenses Against KV-cache Side Channels}}
To mitigate KV-cache side-channel attacks, researchers and practitioners have proposed a range of defenses. 

The most straightforward solution is \textbf{user-level cache isolation}: it prevents cross-user sharing by allocating distinct cache namespaces per user. This containment strategy eliminates cache-based interactions between users and is adopted by some LLM providers (OpenAI~\cite{openai2024promptcaching} and DeepSeek~\cite{deepseek2024promptcaching}) and by researchers~\cite{pang2024cache} for prefix caching. While effective, such strict isolation sacrifices memory efficiency and undermines the performance benefits of cache reuse.

\textbf{Rate Limiting} serves as a complementary defense by throttling the frequency of user queries, thereby impeding rapid probing required for cache-timing attacks. For instance, OpenAI enforces rate limits to prevent abuse and ensure infrastructure stability~\cite{openai2024ratelimits}. However, rate limiting must be carefully tuned to avoid degrading service quality for benign users.

A third line of defense is \textbf{Timing Obfuscation}, which aims to eliminate observable latency differences between cache hits and misses. Prior work in model extraction has shown that response time can correlate with internal architecture details~\cite{batina2019csi,duddu2018stealing,dong2019floating}, motivating similar countermeasures in LLM serving. Two common strategies are: (i) enforcing constant-time execution~\cite{maji2021leaky} or injecting random delays to flatten latency variance~\cite{breier2023desynchronization}, and (ii) disabling token-level streaming, which removes fine-grained timing signals from observable outputs. While these approaches can mask timing patterns, they often incur latency penalties or reduce the interactivity of real-time systems.
\section{Conclusion}
\label{sec:conclusion}

% This paper presents \systemname, a privacy-preserving KV-cache management framework for LLM serving systems, designed to mitigate timing side-channel attacks arising from shared cache entries. \systemname\ combines a hybrid privacy detection pipeline, comprising rule-based matching, lightweight LLM detectors, and context-aware validation, with a sensitivity-aware KV-cache manager that supports efficient reuse through a unified radix tree index.

% Our multi-tiered detection achieves accurate privacy classification with minimal latency, while the cache manager enables fine-grained isolation without compromising performance. Extensive evaluations show that \systemname\ mitigates over 94\% of timing-based attacks across multiple LLM backbones and improves throughput by up to 2.66× compared to per-user cache isolation.These results demonstrate that selective KV-cache sharing, guided by efficient privacy detection, provides a practical balance between security and scalability. 

This paper presented \systemname, a system-level framework for privacy-preserving KV-cache management in LLM serving. The design integrates hybrid privacy detection with cache-level sensitivity control to mitigate timing side channels introduced by shared caches. By combining rule-based filtering, lightweight model classification, and context-aware validation with a unified radix-tree cache manager, \systemname\ enables selective reuse of non-sensitive entries while securely isolating private content. Evaluations across multiple LLMs show that \systemname\ improves throughput by up to $2.66\times$ compared to per-user cache isolation. Our experiments confirm that privacy-aware cache sharing is not only feasible but essential for scalable and secure LLM serving.
%\newpage

%%%%%%%%% -- BIB STYLE AND FILE -- %%%%%%%%
\bibliographystyle{ACM-Reference-Format}
\bibliography{references}
%%%%%%%%%%%%%%%%%%%%%%%%%%%%%%%%%%%%

% \newpage
% \input{appendix}

\end{document}